\documentclass[aps, showpacs, showkeys,nofootinbib,floatfix ,twocolumn]{revtex4}

\usepackage{amssymb}
\usepackage{amsmath}
\usepackage{graphicx}

\begin{document}

\title{Spherical Collapse for Viscous Generalized Chaplygin Gas Model}

\author{Wei Li$^{1,2}$\footnote{liweizhd@126.com}}
\author{Lixin Xu$^{1}$\footnote{corresponding author:lxxu@dlut.edu.cn}}


\affiliation{$^{1}$Institute of Theoretical Physics, School of Physics and Optoelectronic Technology,
Dalian University of Technology, Dalian, 116024, China}
\affiliation{$^{2}$Department of Physics, Bohai University, Jinzhou,
121013, People¡¯s Republic of China}

\begin{abstract}
The nonlinear collapse for viscous generalized Chaplygin gas Model (VGCG) was analyzed in the framework of spherical top-hat collapse. As the VGCG and baryons are essential to form the large scale structure, we focused on their nonlinear collapse in this paper. We discussed the influence of model parameters $\alpha$ and $\zeta_{0}$ on the spherical collapse by varying their values and compare with $\Lambda
CDM$. The results show that, for the VGCG model, smaller $\zeta_{0}$ and larger $\alpha$ make the structure formation earlier and faster, and the collapse curves of VGCG model is almost distinguished with the $\Lambda
CDM$ model when the model parameter $\alpha$ is less than $10^{-2}$.
\end{abstract}



\maketitle

\section{Introduction}

 In the recently years, an increasing number of cosmological observations suggest that our universe is filled with an imperfect fluid which including bulk viscosity in its negative pressure, this pressure was dubbed effective pressure as was argued in \cite{A. B. Balakin}\cite{W. Zimdahl}. Based on this condition, viscous generalized chaplygin gas models \cite{C. S. J. Pun, Wei Li1, Wei Li2, XIANG-HUA ZHAI, A. R. Amani} which being an competitive model to be used to explain the late time accelerated expansion of universe were investigated extensively. In these literatures, the VGCG model unified dark energy and cold dark matter as a unique imperfect dark fluid, which continuing to have retain the property of simulating the $\Lambda$CDM model well on the background level.

Usually, the bulk viscosity is chosen to be a density-dependent or time-dependent function. A density-dependent viscosity $\zeta=\zeta_{0}\rho^{m}$ coefficient is widely investigated in some relevant literature, where $\zeta_{0}>0$ ensures a positive entropy at the request of the second law of thermodynamics. In our previous work \cite{Wei Li1}\cite{Wei Li2}, we studied the case $m=\frac{1}{2}$, and obtained good results in line with the cosmic observations. If a model cannot describe the observed large scale structure and the background evolution, it should be ruled out because of a conflict among the cosmic observations and the theoretical calculation, and VGCG model is no exception.
Because the universe original perturbations are the seed of the large-scale structure, investigating the evolutions of density perturbations of realistic cosmological model become very important. During this process, the study of non-linear perturbations is inevitable. To best of our knowledge, hydrodynamical/N-body numerical simulation (see,
e.g., \cite{A.V. Maccio,N. Aghanim, M. Baldi, B. Li}) is a cumbersome task which is usually used to handle with a fully nonlinear analysis. Fortunately, there is a simple framework to solve this issue.
In \cite{R. A. A. Fernandes}, the no-linear collapse of general Chaplygin gas was investigated in the frame of spherical top-hat collapse, they come to a conclusion that with increasing values of $\alpha$, the growth of the structure becomes faster. In this paper, we expand their work by considering bulk viscosity in the general Chaplygin gas model (VGCG). Besides the parameter $\alpha$, we will also analyze the effect of bulk viscosity $\zeta_{0}$ on the structure formation of the VGCG model which has a spherically symmetric perturbation.

The paper is organized as follows. In section \ref{sec:review}, we give a brief review of the VGCG model and present some basic equations for spherical top-hat collapse. Section \ref{sec:perevolution} is the method and main results. The conclusion is present in the last section.

\section{The Basic Equations for Spherical Top-hat Collapse of VGCG model}\label{sec:review}
In an isotropic and homogeneous universe, the effective pressure of viscous generalized Chaplygin gas (VGCG) model \cite{Wei Li1, Wei Li2} is given in the form \cite{S. Capozziello,Sergei D. Odintsov,I. Brevik}
\begin{equation}
p_{VGCG}=-A/\rho^{\alpha}_{VGCG}-\sqrt{3}\zeta_{0}\rho_{VGCG},
\end{equation}
and the equaton of energy density is
\begin{eqnarray}
\rho_{VGCG}&=&\rho_{GCG0}[\frac{B_{s}}{1-\sqrt{3}\zeta_{0}}+(1-\frac{B_{s}}{1-\sqrt{3}\zeta_{0}})\nonumber
\\&\times&a^{-3(1+\alpha)(1-\sqrt{3}\zeta_{0})}]^{\frac{1}{1+\alpha}},\label{eq:mcg}
\end{eqnarray}
where $B_{s}=A/\rho^{1+\alpha}_{GCG0}$, $\alpha$ and $\zeta_{0}$ are
model parameters and $0\le
B_s \le 1$ and $\zeta_{0}<\frac{1}{\sqrt{3}}$ are demanded. One can obtain the standard $\Lambda$CDM model when $\alpha=0$ and $\zeta_{0}=0$.
By considering VGCG as a unified component and taking the assumption of a purely
adiabatic perturbations, it is easy to get the Friedmann equation
\begin{eqnarray}
H^{2}&=&H^{2}_{0}\{(1-\Omega_{b}-\Omega_{r}-\Omega_{k})[\frac{B_{s}}{1-\sqrt{3}\zeta_{0}}\nonumber\\
&&+(1-\frac{B_{s}}{1-\sqrt{3}\zeta_{0}})a^{-3(1+\alpha)(1-\sqrt{3}\zeta_{0})}]^{\frac{1}{1+\alpha}}\nonumber\\
&&+\left.\Omega_{b}a^{-3}+\Omega_{r}a^{-4}+\Omega_{k}a^{-2}\right\},
\end{eqnarray}
and the effective adiabatic sound speed for VGCG
\begin{equation}
c^2_{ad,eff}=\frac{\dot{p}_{VGCG}}{\dot{\rho}_{VGCG}}=-\alpha
w_{eff}-\sqrt{3}\zeta_{0},\label{eq:cs2}
\end{equation}
where $w_{eff}$ is the EoS of VGCG in the form of
\begin{eqnarray}
w_{eff}&=&w-\sqrt{3}\zeta_{0}\notag\\
&=&-\frac{B_{s}}{B_{s}+(1-B_{s})a^{-3(1+\alpha)}}-\sqrt{3}\zeta_{0},
\end{eqnarray}
on account of the negative values of $w_{eff}$, $\alpha\ge 0$ is required in order to ensure that the sound of speed is non-negative.

The spherical collapse (SC) which provides a way to glimpse into the nonlinear regime of perturbation theory was introduced firstly by Gunn and Gutt 1972 \cite{J. E. Gunn}. Following the assumption of a top-hat profile, namely the density perturbation is uniform throughout the
collapse, so the evolution of perturbation is only time-dependent. That is to say, we can let the gradients inside the perturbed region alone as it is managed in the \cite{R. A. A. Fernandes}.

In the spherical top-hat collapse (SC-TH) model, the equations for background evolution are
\begin{eqnarray}
\dot{\rho}&=&-3H(\rho+p),\\
 \frac{\ddot{a}}{a}&=&-\frac{4\pi G}{3}\sum_i(\rho_i+3p_i),
\end{eqnarray}
and the basic equations in the perturbed region are
\begin{eqnarray}
\dot{\rho}_c&=&-3 h(\rho_c+p_c),\\
 \frac{\ddot{r}}{r}&=&-\frac{4\pi G}{3}\sum_i(\rho_{c_i}+3p_{c_i}),
\end{eqnarray}
where $\rho_c=\rho+\delta\rho$, $p_c=p+\delta
p$ are the perturbed quantities, and $h$ relates to $H$ rate in the STHC framework \cite{R. A. A. Fernandes}\cite{ref:Abramo2009},
\begin{equation}
h=H+\frac{\theta}{3a},
\end{equation}
where $\theta\equiv\nabla\cdot \overrightarrow{v}$ is the divergence of the peculiar velocity $\overrightarrow{v}$.

So, the dynamical evolution equations of density contrast $\delta_i=(\delta\rho/\rho)_i$ and $\theta$ are in the following form
\begin{eqnarray}
\delta'_i&=&-\frac{3}{a}(c^2_{e_i}-w_i)\delta_i-[1+w_i+(1+c^2_{e_i})\delta_i]\frac{\theta}{a^2H},\label{eq:deltaa}\\
\theta'&=&-\frac{\theta}{a}-\frac{\theta^2}{3a^2H}-\frac{3H}{2}\sum_i\Omega_i\delta_i(1+3c^2_{e_i}),\label{eq:thetaa}
\end{eqnarray}
and the equation of state $w_c$ is
\begin{equation}
w_c=\frac{p+\delta p}{\rho+\delta\rho}=\frac{w_{eff}}{1+\delta}+c^2_e\frac{\delta}{1+\delta},
\end{equation}
and the the most important quantity effective sound speed is
\begin{equation}
c^2_{e}=\frac{\delta p}{\delta\rho}=\frac{p_c-p}{\rho_c-\rho}=-\alpha
w_{eff}-\sqrt{3}\zeta_{0}.
\end{equation}
For the details of these equations, please look into the papers \cite{R. A. A. Fernandes} and literatures therein.

\section{The Method and Results}\label{sec:perevolution}

In the following, for studying the non-linear evolution of the baryon and VGCG perturbations in the frame of spherical top-hat collapse, we will perform a mathematical simulation via the software {\bf Mathematica}. In this process, we solve the differential equations (\ref{eq:deltaa})(\ref{eq:thetaa}) through setting the initial conditions (ICs) $\delta_{d}(z=1000)=3.5\times 10^{-3}$ , $\delta_{b}(z=1000)=10^{-5}$, $\delta_{d}$ and $\delta_{b}$ and $\theta=0$ which are the same conditions used in Ref. \cite{R. A. A. Fernandes}\cite{collapse}.

\begin{center}
\begin{figure}[htb]
\includegraphics[width=8cm]{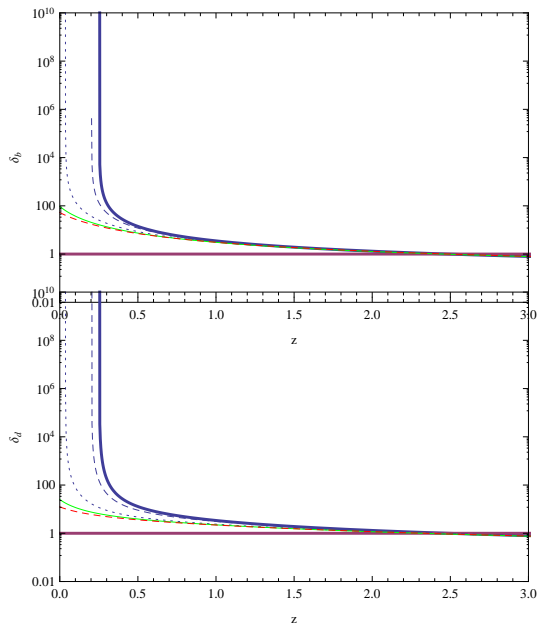}
\caption{The evolutions of density perturbations with respect to the redshift for VGCG models, where the bulk viscosity coefficient is fixed on $\zeta_{0}=0.000708$ and the thick, dashed, dotted and solid curved lines are for $\alpha=1, 0.5, 0.1, 0.01$ respectively. The top and bottom panels are for baryons and VGCG respectively.  The horizon line i.e. $\delta=1$ denotes the limit of linear perturbation and the vertical parts of the curved lines are the collapse of the perturbed regions.}\label{fig:a}
\end{figure}
\end{center}

\begin{center}
\begin{figure}[htb]
\includegraphics[width=8cm]{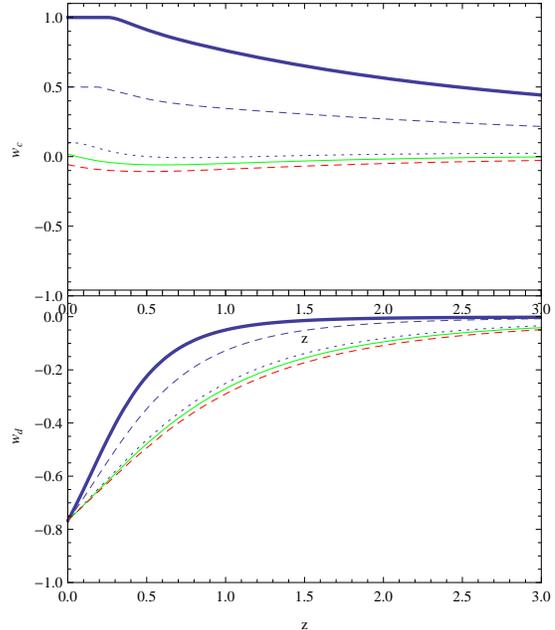}
\caption{The evolutions of $w_c$ and $w_d$ with respect to the redshift $z$ for VGCG models, where the thick, dashed, dotted and solid curved lines are for $\alpha=1, 0.5, 0.1, 0.01$ respectively. The top and bottom panels are for $w_c$ and $w_d$ respectively. }\label{fig:wa}
\end{figure}
\end{center}

In order to show the influence of the model parameter $\alpha$ and $\zeta_{0}$ on the spherical collapse , we let the other relevant cosmological model parameters take their central values $H_0=70.324 \text{km s}^{-1}\text{Mpc}^{-1}$, $\Omega_{d}=0.954$, $\Omega_{b}=0.046$, and $B_s= 0.766$ which obtained in Ref. \cite{Wei Li1}. At first, let's investigate the impact of  parameter $\alpha$ on the non-linear collapse.
By fixing the $\zeta_{0}=0.000708$ which is gotten in our previous work \cite{Wei Li1} and varying the model parameter $\alpha=1$, $0.5$, $0.1$, and $0.01$ respectively, we get the results as shown in Table \ref{tab:alpha} and in Figure \ref{fig:a}, \ref{fig:wa}, where $z_{ta}$ is the turnaround redshift when the collapse of perturbed region is begining. Moreover, we plot the collapse curves of $\Lambda
CDM$ model using the red dashed curves in the two figures above to compare it with the VGCG model. From these results, one can conclude that the perturbations collapse earlier for the larger values of $\alpha$, furthermore, the collapse curves of VGCG model is almost distinguished with the $\Lambda
CDM$ model when the model parameter $\alpha$ is less than $10^{-2}$. This conclusion is the same as the result which obtained in the previous papers, such as the  Ref. \cite{R. A. A. Fernandes}\cite{collapse}.

\begin{table}[tbh]
\begin{center}
\begin{tabular}{ccccc}
\hline\hline Model & $\alpha$ & $\zeta_{0}$ & $B_{s}$ & $z_{ta}$  \\ \hline
a & $0$    & $0$        & $0.766$ & $ 0.104$\\
b & $0.01$ & $0.000708$ & $0.766$ & $ 0.128$\\
c & $0.1$ & $0.000708$ & $0.766$ & $0.251$\\
d & $0.5$ & $0.000708$ & $0.766$ & $0.667$\\
e & $1$ & $0.000708$ & $0.766$ & $0.785$ \\
\hline\hline
\end{tabular}
\caption{Models for the STHC model, where the values of $\alpha$ are small nonnegative values because of the constraint from background evolution. Note that, the model "a" is identical to $\Lambda
CDM$ model. The redshift $z_{ta}$ is the turnaround redshift when the collapse of perturbed region is begining.}\label{tab:alpha}
\end{center}
\end{table}

\begin{center}
\begin{figure}[htb]
\includegraphics[width=8cm]{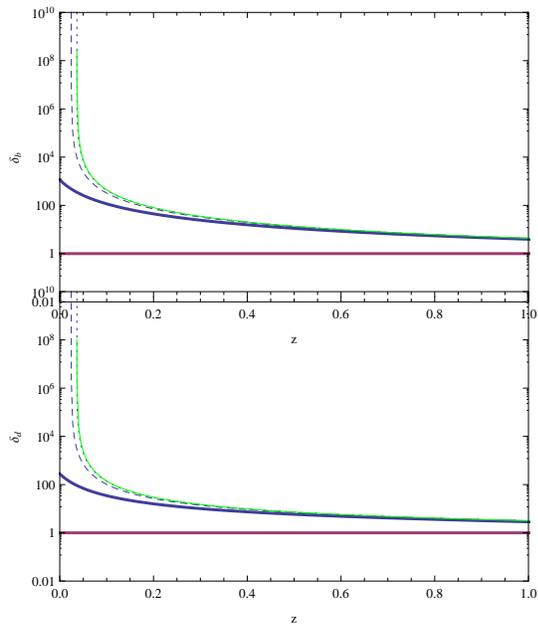}
\caption{The evolutions of density perturbations with respect to the redshift for VGCG models, where the model parameter is fixed on $\alpha=0.035$ and the thick, dashed, dotted and solid curved lines are for $\zeta_{0}=10^{-3}, 10^{-4}, 10^{-5}, 0$ respectively. The top and bottom panels are for baryons and VGCG respectively. The horizon line i.e. $\delta=1$. denotes the limit of linear perturbation and the vertical parts of the curved lines are the collapse of the perturbed regions.}\label{fig:zeta0}
\end{figure}
\end{center}

\begin{center}
\begin{figure}[htb]
\includegraphics[width=8cm]{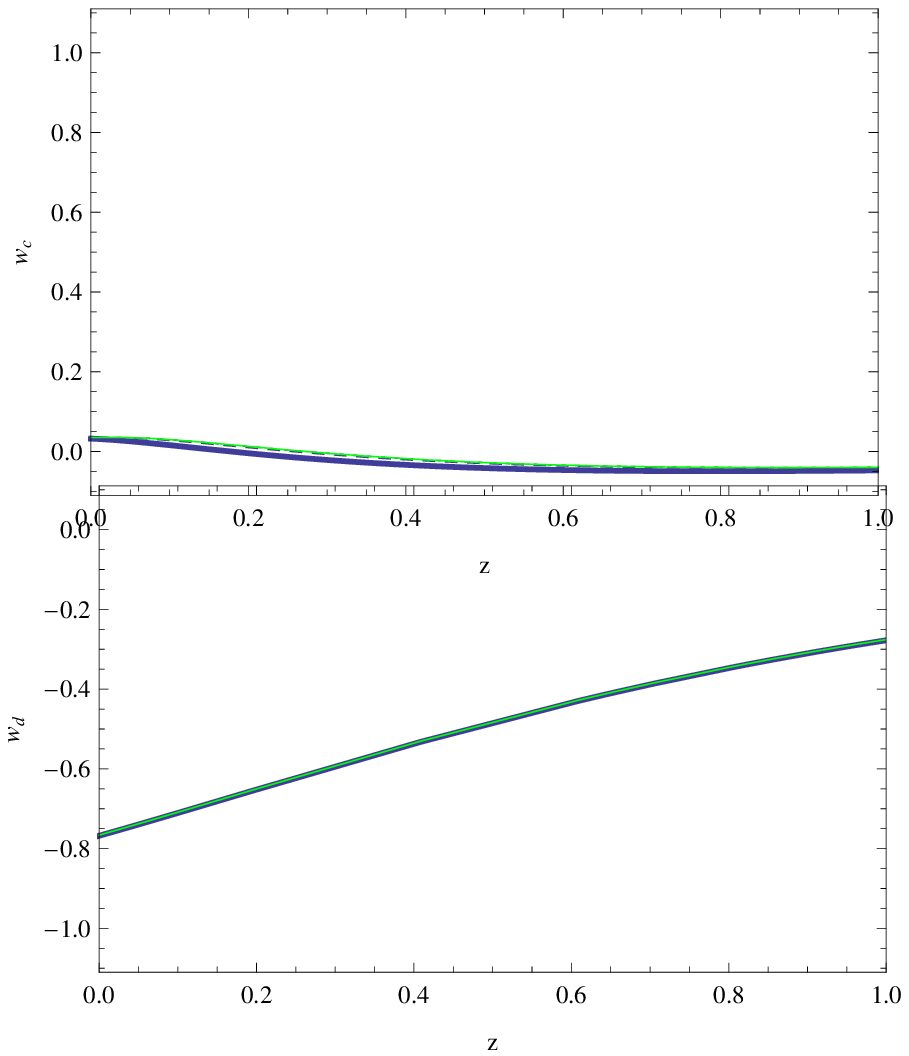}
\caption{The evolutions of $w_c$ and $w_d$ with respect to the redshift $z$ for VGCG model $\alpha=0.035$, where the thick, dashed, dotted and solid curved lines are for $\zeta_{0}=10^{-3}, 10^{-4}, 10^{-5}, 0$ respectively. The top and bottom panels are for $w_c$ and $w_d$ respectively. }\label{fig:w zeta0}
\end{figure}
\end{center}

Next, we will show the effect of $\zeta_{0}$ on the evolution of the density perturbations in the VGCG model. Here we fix $\alpha=0.035$ which is borrowed from our previous work \cite{Wei Li1} and change the values of bulk viscosity $\zeta_{0}=0.001$, $0.0001$, $0.00001$, and $0$ respectively. The corresponding evolutions of density perturbations of  the baryon and VGCG are shown in Figure \ref{fig:zeta0} and the evolutions of EOS parameter are displayed in the Figure \ref{fig:w zeta0}. Seeing from the Figure \ref{fig:zeta0}, the horizon line $\delta=1$ denotes linear perturbation limit and the vertical parts of the curved lines stands for the perturbed regions collapse, so one can conclude that the smaller bulk viscosity coefficient $\zeta_{0}$ can lead to the earlier collapse, that is to say, the larger the value of $\zeta_{0}$ is, the later the collapse comes about. Therefore, this is the reason that the bulk viscosity coefficient $\zeta_{0}$ shouldn't be too large.

Through analysis above, one can clearly comprehend the impact of the model parameters $\zeta_{0}$ and $\alpha$ on the evolutions of the density perturbations. In addition, we can draw a conclusion that the influence of model parameter $\alpha$ is much stronger than the bulk viscosity coefficient $\zeta_{0}$.
The reason is that $\alpha$ is closely linked with the effective sound speed which characters the propagation velocity of the perturbations.

\section{Conclusion} \label{sec:conclusion}

In the present paper, we discussed the structure formation of the viscous generalized Chaplygin gas model in the spherical top-hat collapse framework. We studied the effects of $\zeta_{0}$ and $\alpha$ on the non-linear perturbation evolutions via choosing their different values and compare with $\Lambda
CDM$ model. On the basis of the calculations and analysis, one can come to a conclusion that large $\alpha$ and small $\zeta_{0}$ can bring about a earlier and faster collapse, and when the model parameter $\alpha$ is less than $10^{-2}$, the collapse curves of VGCG model almost overlap with $\Lambda
CDM$ model. Moreover,we can also get the result that the influence of $\alpha$ on the large scale structure formation is remarkable than $\zeta_{0}$. In the next work, we will try to study non-linear collapse by using the hydrodynamical/N-body numerical simulation.

\section{Acknowledgements}

L. Xu's work is supported in part by NSFC under the Grants No. 11275035 and "the Fundamental Research Funds for the Central Universities" under the Grants No. DUT13LK01.

\end{document}